\documentclass[conference]{IEEEtran}
\IEEEoverridecommandlockouts

\usepackage{cite}
\usepackage{amsmath,amssymb,amsfonts}
\usepackage{algorithmic}
\usepackage{graphicx}
\usepackage{textcomp}
\usepackage{xcolor}
\def\BibTeX{{\rm B\kern-.05em{\sc i\kern-.025em b}\kern-.08em
    T\kern-.1667em\lower.7ex\hbox{E}\kern-.125emX}}

\usepackage{booktabs}
\usepackage{siunitx}
\usepackage{tikz}
\usetikzlibrary{positioning}
\usepackage{pgfplots}
\usepackage{listings}
\lstdefinelanguage{json}{
    basicstyle=\ttfamily\small,
    showstringspaces=false,
    breaklines=true,
    literate=
     *{0}{{{\color{black}0}}}{1}
      {1}{{{\color{black}1}}}{1}
      {2}{{{\color{black}2}}}{1}
      {3}{{{\color{black}3}}}{1}
      {4}{{{\color{black}4}}}{1}
      {5}{{{\color{black}5}}}{1}
      {6}{{{\color{black}6}}}{1}
      {7}{{{\color{black}7}}}{1}
      {8}{{{\color{black}8}}}{1}
      {9}{{{\color{black}9}}}{1}
}
\lstdefinelanguage{yaml}{
    basicstyle=\ttfamily\small,
    showstringspaces=false,
    breaklines=true
}
\pgfplotsset{compat=1.18}

\newcommand{\cai}{\ensuremath{\mathrm{CAI}}}
\newcommand{\oas}{\ensuremath{\mathrm{OAS}}}
\newcommand{\psw}{\ensuremath{\mathrm{PSW}}}
\newcommand{\iil}{\ensuremath{\mathrm{IIL}}}
\newcommand{\ei}{\ensuremath{\mathrm{EI}}}
\newcommand{\chr}{\ensuremath{\mathrm{CHR}}}
\newcommand{\crg}{\ensuremath{\mathrm{CRG}}}
\newcommand{\css}{\ensuremath{\mathrm{CSS}}}
\newcommand{\fcr}{\ensuremath{\mathrm{FCR}}}

\title{Policy-Governed LLM Routing with Intent Matching for Instrument Laboratories}

\author{\IEEEauthorblockN{Emmanuel A. Olowe}
\IEEEauthorblockA{\textit{School of Engineering} \\
\textit{The University of Edinburgh}\\
Edinburgh, UK \\
e.a.olowe@sms.ed.ac.uk}
\and
\IEEEauthorblockN{Danial Chitnis}
\IEEEauthorblockA{\textit{School of Engineering} \\
\textit{The University of Edinburgh}\\
Edinburgh, UK \\
d.chitnis@ed.ac.uk}
}

\begin{document}
\maketitle

\begin{abstract}
AI tutoring systems in engineering labs face a tension between providing sufficient assistance and preserving learning opportunities. Existing systems typically offer instructors limited control over assistance timing, content, or cost. This paper describes a routing and governance system for LLM-based lab assistance comprising two components: Routiium, an OpenAI-compatible gateway that manages multiple LLM backends with configurable prompt modifications and usage logging, and EduRouter, a policy-aware routing service that enforces per-lab budgets, approval workflows, and embedding-based question matching.

We evaluated the system using trace-driven simulation calibrated from two engineering labs (LED characterization, RC circuit analysis) and a 100-query replay through live models. In simulations, governed policies (P1/P2) increased challenge-alignment index from 0.90 to 0.98 and overlay-adherence score from 0.69 to 0.87 compared to ungoverned operation (P0). The productive-struggle window metric increased from 1.4 to 3.6 simulated turns before high-scaffold hints appeared. In the 100-query replay, EduRouter routed 75\% of queries to a local model, reducing token costs by 66\% (\$0.087 vs.\ \$0.26 for all-premium routing) while maintaining canonical hit rate of 1.0 for the curated 89-intent question bank. We release Routiium, EduRouter, canonical-task tooling, and simulator configurations to support replication and future classroom studies.
\end{abstract}

\begin{IEEEkeywords}
AI in education, routing, prompt control, remote laboratories, steerability, simulation
\end{IEEEkeywords}

\section{Introduction}

Large language models (LLMs) are increasingly used as ad hoc tutors in engineering courses, including instrument-based laboratory work. In these settings, the same student question can be answered in ways that differ not only in correctness but also in pedagogical consequence: a minimal check that preserves student agency, a guided troubleshooting step, a worked-example fragment, or a complete solution. Classic tutoring and instructional science has long treated this as a control problem: how to balance information giving and withholding so that assistance supports learning without inducing dependence. Early work that introduced the notion of instructional scaffolding emphasized contingent support that is adjusted to learner need~\cite{wood1976role}. In intelligent tutoring systems (ITS), this tension is formalized as the \emph{assistance dilemma}, where different help policies can change student engagement and learning outcomes, with no single policy dominating across contexts~\cite{koedinger2007exploring,vanlehn2011relative}.

Current LLM tutoring deployments \cite{maurya2025unifying, scarlatos2025training} typically provide instructors limited control over (i) when higher-scaffold help is allowed, (ii) what transformations are applied to prompts and responses,  and (iii) how usage costs are bounded and audited. These limitations are operational as well as pedagogical: without explicit budgets, approval workflows, and transparency mechanisms, instructors cannot reliably enforce course policies (for example, restricting complete solutions during assessed steps) or explain system behavior when students challenge a decision. Recent reviews of LLMs in education describe both potential benefits and risks, including concerns about over-reliance and integrity, but do not prescribe implementable control mechanisms for lab-style tutoring at turn level~\cite{kasneci2023chatgpt, mollick2023using}.

In parallel, systems research on LLM routing has focused on reducing cost or latency by selecting among multiple models (for example via cascades or learned routers)~\cite{chen2023frugalgpt,ong2024routellm,huang2025thriftllm}. However, these routers typically optimize for answer quality proxies and budget constraints, not for instructor-defined help policies such as ``permit only conceptual guidance unless approved'' or ``cap complete-solution frequency per student.'' Separately, retrieval and embedding methods provide a practical mechanism for mapping student questions to recurring intents (for example via sentence embeddings and dense retrieval)~\cite{reimers2019sentence,gao2021simcse,karpukhin2020dense}. Yet existing educational deployments rarely connect embedding-based intent matching to auditable governance actions (approval queues, spend debits, policy overlays) in instrumented lab environments.

This paper addresses the following gap: \emph{there is limited systems evidence on how to combine cost-aware multi-model routing with instructor-governed scaffolding policies and auditable prompt interventions for instrument laboratories}. We describe a routing and governance system comprising two components. Routiium is an OpenAI-compatible gateway that fronts multiple LLM backends, applies configurable prompt modifications (overlays), and logs per-turn usage. EduRouter is a policy-aware routing service that enforces per-lab budgets, approval workflows, and embedding-based question matching against an instructor-curated library of common lab intents.

Our design uses embedding similarity to match student queries to a canonical question library. When a match exceeds a configurable threshold, routing consults the matched entry's metadata (preferred model tier, permitted hint level, maximum cost, overlay choice). When no match is available, routing falls back to heuristic selection. The goal is not to infer pedagogy from first principles, but to implement instructor-specified rules with traceable, testable behavior.

We evaluate the system in two ways. First, we use trace-driven simulation calibrated from two electronics labs to measure whether routing and governance actions conform to specified policies under synthetic workloads. Second, we replay 100 curated queries through live model endpoints to measure cost, latency, and routing correctness relative to the canonical library. These evaluations characterize system behavior under the stated conditions; they do not measure learning outcomes, instructor workload reduction, or student satisfaction. Consistent with this scope, we treat our metrics as behavioral compliance measures rather than educational effectiveness measures~\cite{siemens2013learning,pardo2014ethical}. Section~\ref{sec:threats} details threats to validity, including the constructed nature of the replay workload and the limited calibration data.

This paper makes the following contributions:
\begin{itemize}
    \item Routiium, an OpenAI-compatible gateway enforcing configurable prompt overlays with per-turn logging and auditability.
    \item EduRouter, a routing service implementing policy-aware model selection with embedding-based matching to an instructor-curated question library.
    \item A set of governance-oriented metrics that characterize alignment between observed help behavior and instructor-specified policies, without asserting educational efficacy.
    \item Synthetic and replay-based evaluations that quantify policy compliance, cost, and latency under the stated experimental constraints.
    \item Open-source release of all components and evaluation tooling.
\end{itemize}

\section{Background and Related Work}

\subsection{Scaffolding, the Assistance Dilemma, and Help-Seeking}
Work on instructional scaffolding emphasizes contingent support that is adjusted to learner need rather than uniformly increasing explanation detail~\cite{wood1976role}. In ITS research, this is operationalized as the assistance dilemma: policies that provide more help can improve short-term task completion while risking reduced learner engagement, and the conditions under which help is beneficial remain context dependent~\cite{koedinger2007exploring}. Comparative reviews of tutoring systems report substantial variance by domain and implementation, reinforcing that system designers should avoid assuming a single optimal help policy~\cite{vanlehn2011relative}. Our system does not propose a new theory of when help should be given; it implements instructor-specified help policies (hint levels, approvals, frequency caps) and provides instrumentation to audit compliance.

\subsection{Teacher Agency and Classroom Orchestration}
Prior work on classroom orchestration treats teacher intervention as a first-class systems requirement, where tools should support monitoring and timely control actions rather than only automate tutoring~\cite{dillenbourg2011classroom}. Recent studies on teacher--AI complementarity similarly argue for interfaces and workflows that let teachers shape and audit automated support~\cite{holstein2019co}. We operationalize this perspective for instrument labs by providing turn-level governance actions (approval queues, budgets, freezes) plus audit logs that expose routing rationale and applied prompt overlays.

\subsection{Cost-Aware LLM Routing}
Several systems route queries across multiple LLM backends to reduce costs. FrugalGPT~\cite{chen2023frugalgpt} cascades models from cheapest to most expensive, stopping when output confidence exceeds a threshold. RouteLLM~\cite{ong2024routellm} learns routing policies from preference data. ThriftLLM~\cite{huang2025thriftllm} optimizes selection under budget constraints. Pulishetty et al.~\cite{pulishetty2025one} explore cross-attention mechanisms for routing.

These systems optimize for answer quality or cost, not pedagogical appropriateness. A response that is optimal by quality metrics may provide more detail than an instructor intends for a given learning stage. We adapt cost-aware routing to include policy constraints on hint levels and approval requirements, though whether these constraints improve educational outcomes remains untested.

Embedding-based matching is a standard approach for mapping semantically similar queries to shared intents in retrieval settings~\cite{reimers2019sentence,gao2021simcse,karpukhin2020dense}; our canonical bank plays a similar role, but its outputs are consumed by governance policies (hint limits, approvals, budget debits) rather than only by a generation-time context augmenter~\cite{lewis2020retrieval}.

\subsection{Remote Labs and Teacher Dashboards}
Remote labs enable equipment access over networks~\cite{veza2022virtual, van2023remote, poo2023virtual}. Platforms including ReImagine Lab~\cite{9858137} and LabEAD~\cite{s22186944} demonstrate scalable architectures.

\subsection{AI Assistance in Remote Laboratories}
Remote laboratories are widely used to provide networked access to physical instruments, and some systems have begun integrating intelligent assistance into these environments~\cite{rassudov2022virtual}. However, these efforts do not typically foreground cost-aware multi-model routing coupled to instructor-governed help policies with auditable prompt interventions. Our contribution is therefore not ``AI in remote labs'' per se, but a governance and routing layer that can be evaluated as a systems artifact and attached to existing remote-lab platforms.

Teacher dashboards have been developed for monitoring and intervention~\cite{verbert2020learning,karademir2024following, han2021learning, rienties2018making}, with recent work examining alternatives to traditional dashboard designs~\cite{fernandez2022beyond}. Research on system prompts~\cite{polignano2024unraveling,neumann2025position,wang2024rnr} has raised concerns about hidden prompt modifications affecting outputs. Our system logs all prompt modifications and makes them visible to instructors through the audit interface.

\begin{table*}[t]
\caption{Positioning relative to related systems (illustrative dimensions).}
\label{tab:positioning}
\centering
\begin{tabular}{@{}lcccc@{}}
\toprule
System & Multi-model routing & Budget enforcement & Instructor approvals & Prompt/audit transparency \\
\midrule
FrugalGPT~\cite{chen2023frugalgpt} & \checkmark & \checkmark & --- & --- \\
RouteLLM~\cite{ong2024routellm} & \checkmark & (varies) & --- & --- \\
ThriftLLM~\cite{huang2025thriftllm} & \checkmark & \checkmark & --- & --- \\

This work (Routiium+EduRouter) & \checkmark & \checkmark & \checkmark & \checkmark \\
\bottomrule
\end{tabular}
\end{table*}

\subsection{Synthesis and Gap Statement}
Prior work provides (i) pedagogical framing for why assistance timing and granularity matter, including the assistance dilemma~\cite{koedinger2007exploring,vanlehn2011relative}; (ii) systems methods for reducing LLM cost via routing~\cite{chen2023frugalgpt,ong2024routellm,huang2025thriftllm}; and (iii) retrieval and embedding techniques for intent matching~\cite{reimers2019sentence,gao2021simcse}. What is less developed is systems evidence on how to combine these threads in instrument laboratories: implementing instructor-defined help policies (including approvals and hint caps), enforcing budgets, and exposing auditable traces of routing and prompt modifications. The system in this paper is designed to fill this integration gap and is evaluated as a governance-and-routing artifact rather than as a claim about learning effectiveness.

\section{System Overview}

The system integrates with LABIIUM, a platform for remote electronics laboratories~\cite{Olowe2025LABIIUM}. Students interact with instruments through web interfaces; our components add AI assistance to this infrastructure.

The system comprises three parts (Figure~\ref{fig:system}): an instructor control panel for setting policies and reviewing logs, the Routiium gateway managing LLM access, and the EduRouter decision engine selecting backends.

Table~\ref{tab:hintpolicy} defines the hint levels (L0--L3) and policy modes (P0--P2) referenced throughout the paper.

\begin{table}[t]
    \caption{Hint levels and policy modes referenced throughout the paper.}
    \label{tab:hintpolicy}
    \centering
    \begin{tabular}{@{}p{0.12\columnwidth}p{0.36\columnwidth}p{0.42\columnwidth}@{}}
        \toprule
        Label & Definition & Governance behavior \\
        \midrule
        L0 & Validation or minimal prompt & No approval required; conceptual framing \\
        L1 & Guided troubleshooting hint & Overlay adds justification prompts \\
        L2 & Worked example fragment & Requires budget debit; throttled after configurable retry count \\
        L3 & Complete solution & TA approval required plus budget and frequency limits \\
        \midrule
        P0 & Ungoverned & No spend limits, overlays disabled, no approvals \\
        P1 & Governed & Per-turn spend limits, overlays enabled, L3 frequency caps \\
        P2 & Integrity-focused & Approval queues, stricter overlays, integrity flag gating \\
        \bottomrule
    \end{tabular}
\end{table}

\begin{figure*}[t]
\centering
\begin{tikzpicture}[
    font=\small,
    box/.style={draw, rounded corners, minimum height=8mm, fill=black!3},
    server/.style={draw, rounded corners, thick, fill=blue!10},
    proxy/.style={draw, rounded corners, thick, fill=blue!5},
    lamb/.style={draw, rounded corners, fill=green!8, align=center},
    student/.style={draw, circle, minimum size=8mm, fill=orange!15},
    teacher/.style={draw, rounded corners, thick, fill=red!10},
    device/.style={draw, rounded corners, fill=gray!20, align=center}
]

\node[teacher, text width=12cm, align=left] (teacher) at (0,0) {
    \textbf{Teacher Control Console}\\[2mm]
    \footnotesize
    $\triangleright$ Monitor all 30 LAMBs \quad $\triangleright$ Set hint levels (L0--L3) \quad $\triangleright$ Cost \& budget control\\
    $\triangleright$ Approval queue \quad $\triangleright$ Add/modify preconditions \quad $\triangleright$ Policy updates (P0/P1/P2)\\
    $\triangleright$ Real-time telemetry \quad $\triangleright$ Dynamic config \quad $\triangleright$ One-click boost/throttle
};

\node[server, text width=5.3cm, align=left] (edurouter) at (-4,-3.6) {
    \textbf{EduRouter}\\[1mm]
    \footnotesize
    \textbullet\ Cost- \& policy-aware planner\\
    \textbullet\ Canonical task matching\\
    \textbullet\ Budgets, approvals, freeze TTLs\\
    \textbullet\ Steerability metrics \& logs
};

\node[proxy, text width=5.3cm, align=left] (routiium) at (4,-3.6) {
    \textbf{Routiium proxy}\\[1mm]
    \footnotesize
    \textbullet\ OpenAI-compatible API surface\\
    \textbullet\ Overlay prompts \& tools\\
    \textbullet\ Streaming, TTFT telemetry\\
    \textbullet\ Per-request routing headers
};

\node[font=\scriptsize\bfseries] at (0,-4.75) {LABIIUM routing \& assistance plane};

\node[align=center, font=\footnotesize\bfseries] (lamb_label) at (0,-6.4) {LAMB Devices};

\node[lamb, minimum width=22mm, minimum height=32mm] (lamb1) at (-4.5,-8.5) {};
\node[font=\footnotesize\bfseries] at (-4.5,-7.2) {LAMB 1};
\node[text width=20mm, align=center, font=\scriptsize] at (-4.5,-8.7) {
    VS Code Server\\
    Python Env\\
    AI Executor\\
    Instrument Server\\
    (Rust/VISA)
};

\node[lamb, minimum width=22mm, minimum height=32mm] (lamb2) at (0,-8.5) {};
\node[font=\footnotesize\bfseries] at (0,-7.2) {LAMB 2};
\node[text width=20mm, align=center, font=\scriptsize] at (0,-8.7) {
    VS Code Server\\
    Python Env\\
    AI Executor\\
    Instrument Server\\
    (Rust/VISA)
};

\node[lamb, minimum width=22mm, minimum height=32mm] (lamb30) at (4.5,-8.5) {};
\node[font=\footnotesize\bfseries] at (4.5,-7.2) {LAMB 30};
\node[text width=20mm, align=center, font=\scriptsize] at (4.5,-8.7) {
    VS Code Server\\
    Python Env\\
    AI Executor\\
    Instrument Server\\
    (Rust/VISA)
};

\node[font=\Large] at (2.25,-8.5) {$\cdots$};

\node[device, minimum width=22mm, minimum height=8mm] (inst1) at (-4.5,-11) {Instruments};
\node[device, minimum width=22mm, minimum height=8mm] (inst2) at (0,-11) {Instruments};
\node[device, minimum width=22mm, minimum height=8mm] (inst30) at (4.5,-11) {Instruments};

\node[device, minimum width=22mm, minimum height=8mm] (dut1) at (-4.5,-12.5) {DUT};
\node[device, minimum width=22mm, minimum height=8mm] (dut2) at (0,-12.5) {DUT};
\node[device, minimum width=22mm, minimum height=8mm] (dut30) at (4.5,-12.5) {DUT};

\node[student, font=\scriptsize] (student1) at (-7.5,-8.5) {Student 1};
\node[student, font=\scriptsize] (student2) at (-2.5,-8.5) {Student 2};
\node[student, font=\scriptsize] (student30) at (7.5,-8.5) {Student 30};

\draw[->, very thick, red!60] (teacher.south) -- ++(0,-0.8) -| (edurouter.north) node[pos=0.8, left, font=\scriptsize] {policies \& metrics};
\draw[->, very thick, red!40] (teacher.south) -- ++(0,-0.8) -| (routiium.north) node[pos=0.8, right, font=\scriptsize] {overlays};

\draw[->, thick, blue!70] (edurouter.east) -- (routiium.west) node[midway, above, font=\scriptsize] {\texttt{/route/plan}};

\draw[->, thick, blue!60] (routiium.south) -- ++(0,-0.4) coordinate (lamb1stem) -| (lamb1.north);
\draw[->, thick, blue!60] (lamb1stem) -- ++(0,0) -| (lamb2.north) node[pos=0.6, right, font=\tiny] {HTTPS/QUIC};
\draw[->, thick, blue!60] (routiium.south) -- ++(0,-0.4) -| (lamb30.north);

\draw[<->, thick] (lamb1.south) -- (inst1.north) node[midway, right, font=\tiny] {VISA};
\draw[<->, thick] (lamb2.south) -- (inst2.north) node[midway, right, font=\tiny] {VISA};
\draw[<->, thick] (lamb30.south) -- (inst30.north) node[midway, right, font=\tiny] {VISA};

\draw[->, thick] (inst1.south) -- (dut1.north);
\draw[->, thick] (inst2.south) -- (dut2.north);
\draw[->, thick] (inst30.south) -- (dut30.north);

\draw[->, dashed, orange!70, thick] (student1.east) -- (lamb1.west) node[midway, above, font=\tiny] {web};
\draw[->, dashed, orange!70, thick] (student2.east) -- (lamb2.west) node[midway, above, font=\tiny] {web};
\draw[->, dashed, orange!70, thick] (student30.west) -- (lamb30.east) node[midway, above, font=\tiny] {web};

\draw[dashed, gray!70, thick] (-8.8,1.0) rectangle (9.3,-13.3);
\node[font=\scriptsize, anchor=north east] at (9.3,1.0) {Campus / lab network};

\node[text width=11cm, align=center, font=\scriptsize] at (0,-14) {
    \textbf{Telemetry:} \texttt{tokens, latency, hint\_type, teacher\_boost, scpi\_retries, range\_changes, interlock, step\_pass, rubric\_score}
};

\end{tikzpicture}
\caption{LABIIUM educational platform architecture with centralized teacher control, an explicit routing stack (EduRouter + Routiium), and distributed LAMB devices. The teacher console configures policies, approvals, and overlays; EduRouter computes cost- and policy-aware plans (\texttt{/route/plan}) while Routiium exposes an OpenAI-compatible proxy with overlays, logging, and streaming telemetry. Students reach their assigned LAMB via secure web tunnels, which in turn talk to instruments and DUTs while upstream assistance is routed through this governable routing and assistance plane.}
\label{fig:system}
\vspace{-0.05in}
\end{figure*}

\begin{figure}[t]
\centering
\begin{tikzpicture}[
    font=\scriptsize,
    scale=0.9,
    every node/.style={transform shape},
    proc/.style={draw, rounded corners, thick, fill=blue!8, minimum width=22mm, minimum height=7mm, align=center},
    user/.style={draw, circle, fill=orange!15, minimum size=7mm},
    teacher/.style={draw, rounded corners, thick, fill=red!10, align=center}
]

\node[user] (student) at (-0.8,-0.7) {Student};
\node[teacher, text width=15mm] (teacher) at (-0.2,1.8) {Teacher\\console};

\node[proc] (routiium) at (2.2,0.85) {Routiium\\proxy};
\node[proc] (edurouter) at (5.8,0.85) {EduRouter};
\node[proc] (vllm) at (2.2,-0.9) {vLLM\\(GH200 Server)};
\node[proc, dashed] (openai) at (5.8,-2.05) {OpenAI Servers};

\draw[->, thick, orange!70] (student.east) -- node[above, sloped] {\texttt{/chat}} (routiium.west);

\coordinate (ctrlR) at (routiium.north |- teacher.east);
\coordinate (ctrlE) at (edurouter.north |- teacher.east);
\draw[thick, red!50] (teacher.east) -- (ctrlE);
\draw[->, thick, red!60] (ctrlR) -- node[left, pos=0.5] {overlays/tools} (routiium.north);
\draw[->, thick, red!40] (ctrlE) -- node[left, pos=0.5] {policies/budgets} (edurouter.north);

\draw[->, thick, blue!70] (routiium.east) -- node[above, yshift=0.1em, pos=0.5] {{\tiny\texttt{/route/plan}}} (edurouter.west);

\draw[->, thick, blue!50] (edurouter.west)
    .. controls +(0,-0.5) and +(0,-0.5) ..
    node[below, pos=0.35, xshift=1.2cm, yshift=-0.2em] {\footnotesize plan (tier, model, budget)}
    (routiium.east);

\draw[->, thick, dashed, cyan!60] (routiium.south) -- node[left] {GPT OSS 20b} (vllm.north);
\draw[->, thick, dashed, cyan!60] (routiium.south east)
    .. controls +(0.6,-0.8) and +(-0.6,0.8) ..
    node[right, pos=0.55, yshift=-0.1em] {GPT-5-mini LLM}
    (openai.north west);

\draw[dashed, gray!70, thick] (-1.5,2.3) rectangle (7.0,-1.4);
\node[font=\scriptsize, anchor=north east] at (7.0,2.3) {Campus / lab network};

\end{tikzpicture}
\caption{Protocol-level interaction between student, teacher console, Routiium, EduRouter, and LLM backends. Routiium receives chat requests, consults EduRouter's \texttt{/route/plan}, dispatches to local vLLM or external OpenAI GPT-5 Mini, and streams the response back with routing headers.}
\label{fig:protocol}
\vspace{-0.05in}
\end{figure}
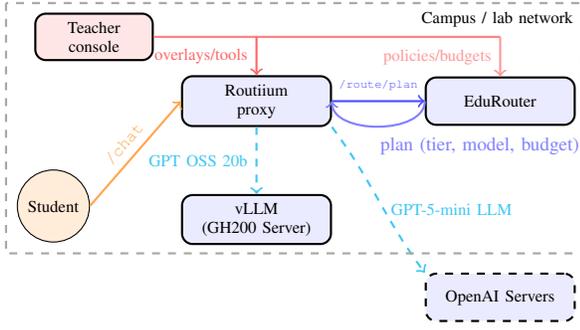

\paragraph*{Usage scenario.} A student asks for wiring confirmation (L0), then requests help fitting a noisy trace (L2), then asks for a complete solution (L3). Under P0, all queries route to the local tier with detailed responses and no approvals. Under P2, the sequence triggers budget debits, conceptual-framing overlays, and TA approval before the L3 response. Instructors see the canonical intent classification, routing decision, cost, and policy compliance status for each turn.

\paragraph*{Instructor console.} The console displays: (1) per-student budget status and pending approvals; (2) turn-level routing metadata including \texttt{X-Route-Why}, \texttt{X-Canonical-Ids}, and cost; (3) controls for adjusting budgets, freezing students on specific models, or modifying overlay settings.

\section{The Routiium Gateway}

Routiium provides five functions: (1) a unified OpenAI-compatible interface to multiple LLM backends; (2) authentication of authorized clients; (3) application of instructor-defined prompt modifications (overlays) to queries and responses; (4) per-turn logging of student ID, model, tokens, cost, and active policy; (5) runtime policy updates without service restart.

The gateway centralizes policy enforcement and provides a privacy boundary: student queries remain within the campus network when routed to local models.

\section{The EduRouter Decision Engine}

EduRouter selects which backend handles each query through four steps: (1) matching the query against the question library using embedding similarity (Section~\ref{sec:embedding-aware-routing}); (2) checking policy constraints and budget state; (3) selecting a model based on library recommendations, cost, and consistency requirements; (4) logging the decision rationale for instructor audit (Table~\ref{tab:headers}).

Consistency (``stickiness'') keeps follow-up queries on the same model for conversational coherence, with configurable TTL. Instructors can manually freeze students on specific models or boost queries to premium models.

\begin{table}[t]
    \caption{Metadata logged for each routing decision.}
    \label{tab:headers}
    \centering
    \begin{tabular}{@{}p{0.31\columnwidth}p{0.63\columnwidth}@{}}
        \toprule
        Field & Content \\
        \midrule
        Routing rationale & Reason for model selection (e.g., ``canonical match: RC circuit troubleshooting'' or ``budget limit enforced'') \\
        Question match & Matched library entry IDs and similarity scores \\
        Active overlay & Which prompt modifications were applied \\
        Content fingerprint & Hash for verifying overlay application \\
        Request trace ID & Link to equipment telemetry \\
        \bottomrule
    \end{tabular}
\end{table}

\section{The Question Library}
\label{sec:embedding-aware-routing}

The question library contains instructor-curated entries representing common lab queries. Each entry includes a text description, preferred model, appropriate hint level, maximum cost, and tags.

Instructors build the library by reviewing historical queries from previous terms and identifying recurring patterns. For an electronics lab, entries might include ``How do I measure rise time on the oscilloscope?'' (basic measurement, L0--L1), ``Why is my circuit oscillating?'' (troubleshooting, L1--L2), or ``Can you explain the relationship between capacitance and time constant?'' (conceptual, L1).

Listing~\ref{lst:canonical} shows an example entry structure.

\begin{lstlisting}[language=json,caption={Example question library entry.},label={lst:canonical}]
{
  "id": "scope_measure_rise_time",
  "text": "Help me set up cursors or automatic measurements to capture rise time and fall time of a digital signal.",
  "preferred_model": "openai/gpt-oss-20b",
  "tags": ["oscilloscope", "basic", "measurement"],
  "overlay": "socratic_troubleshoot",
  "max_cost_usd": 0.05,
  "embedding": {"provider": "bge-large-en-v1.5", "vector_hash": "b9e21a...a44"}
}
\end{lstlisting}

\subsection{Query Matching}
When a student submits a query, the system computes embedding similarity between the query and each library entry, returning entries above a configurable threshold ($\tau = 0.82$ in our experiments). When matches are found, routing uses the matched entry's metadata. When no match exceeds the threshold, the system logs \texttt{X-Route-Why=canonical:none} and falls back to heuristic routing based on query length and estimated complexity.

Matching latency in our deployment averaged 4.7~ms for the 89-entry library.

\subsection{Routing from Matches}
Library matches inform but do not fully determine routing. The system still enforces budget limits, policy constraints, and stickiness requirements. Routing logs include match information so instructors can audit whether library recommendations were followed or overridden by other constraints.

\subsection{Instructor Workflow}
Instructors typically curate a library once per term for each lab, then adjust policies (budgets, approvals, overlays) between sessions based on observed behavior. The authoring process involves exporting anonymized historical queries, clustering by intent, and editing resulting templates. In our experience, creating an 80--100 entry library for a lab takes approximately 2--3 hours.

When queries fall outside library coverage, the router surfaces this in logs, and instructors can add entries for the next term. This workflow assumes instructors are willing to invest curation time; we have not yet studied whether this investment is perceived as worthwhile.

\section{Steerability Metrics}
\label{sec:teacher-metrics}
We define metrics to characterize whether router behavior aligns with instructor-specified policies. These metrics are computed from telemetry logs (Table~\ref{tab:telemetry}) and are intended as behavioral characterizations, not as proxies for educational outcomes. Where feasible, telemetry fields can be mapped to learning analytics event schemas such as Caliper or xAPI, though we do not claim conformance in the current prototype.

Let $\mathcal{S}$ denote lab steps, $\mathcal{T}$ routed turns, $h_s$ the target hint distribution for step $s$, and $\hat{h}_s$ the observed distribution. Teacher actions are drawn from set $\mathcal{A}$ with timestamps $t(a)$, and downstream responses referencing those actions have timestamps $t'(a)$.

\textbf{Challenge Alignment Index (\cai):} Measures agreement between observed hint-level distribution and instructor-specified target distribution.
\begin{equation}
    \cai = 1 - \frac{1}{|\mathcal{S}|} \sum_{s \in \mathcal{S}} \frac{1}{2}\lVert \hat{h}_s - h_s \rVert_1
\end{equation}
A value of 1.0 indicates the observed distribution matched the target exactly.

\textbf{Overlay Adherence Score (\oas):} Fraction of turns where overlay application passed verification.
\begin{equation}
    \oas = \frac{1}{|\mathcal{T}|} \sum_{t \in \mathcal{T}} \mathbf{1}[\texttt{overlay\_guardrail}(t)=\texttt{pass}]
\end{equation}

\textbf{Productive-Struggle Window (\psw):} Average number of turns before a high-scaffold hint (L2 or L3) appears, normalized by step difficulty.
\begin{equation}
    \psw = \frac{1}{|\mathcal{S}|} \sum_{s \in \mathcal{S}} \frac{1}{d_s}\left(\min \{k : \texttt{hint}_k \in \{L2,L3\}\}\right)
\end{equation}
where $d_s \in \{1,2,3\}$ is a difficulty weight from the lab descriptor. Higher values indicate more turns before detailed assistance.

\textbf{Instructor-Influence Latency (\iil):} Median turn delay between instructor action and downstream effect.
\begin{equation}
    \iil = \text{median}_{a \in \mathcal{A}} (t'(a)-t(a))
\end{equation}

\textbf{Equity Index (\ei):} Measures distribution evenness of L3 hints across cohorts.
\begin{equation}
    \ei = 1 - \text{Gini}\left(\{\texttt{L3\_count}(c)\}_{c \in \text{cohorts}}\right)
\end{equation}
A value of 1.0 indicates L3 hints were distributed equally; lower values indicate concentration.

\textbf{Canonical metrics} characterize embedding-aware routing specifically:

\textbf{Canonical Hit Rate (\chr):} Fraction of turns with a library match above threshold $\tau$.
\begin{equation}
    \chr = \frac{1}{|\mathcal{T}|} \sum_{t \in \mathcal{T}} \mathbf{1}[\texttt{score}(t) \ge \tau]
\end{equation}

\textbf{Canonical Routing Gain (\crg):} Cost reduction attributable to embedding routing compared to heuristic baseline, with \cai\ and \oas\ held constant.
\begin{equation}
    \crg = \frac{C_{\text{heur}} - C_{\text{embed}}}{C_{\text{heur}}}
\end{equation}

\textbf{Canonical Stickiness Stability (\css):} Fraction of consecutive turns on the same canonical ID that remained on the same model tier.
\begin{equation}
    \css = \frac{1}{|\mathcal{T}_{\text{canon}}|} \sum_{t \in \mathcal{T}_{\text{canon}}} \mathbf{1}[\texttt{tier}(t)=\texttt{tier}(t-1)]
\end{equation}

\textbf{False Canonical Rate (\fcr):} Fraction of turns with a canonical match that still required fallback routing due to policy or capability constraints.
\begin{equation}
    \fcr = \frac{1}{|\mathcal{T}|} \sum_{t \in \mathcal{T}} \mathbf{1}[\texttt{fallback}(t)=\texttt{true} \land \texttt{score}(t) \ge \tau]
\end{equation}

These metrics characterize system behavior, not educational effectiveness. Whether improvements in these metrics correspond to better learning outcomes is an empirical question we have not addressed.

\section{Evaluation Methodology}
We conducted two synthetic evaluations: a trace-driven simulator computing steerability metrics, and a 100-query replay measuring cost, latency, and routing correctness on live models.

\subsection{Trace-Driven Simulator}
The simulator replays synthetic student workloads through EduRouter, computing metrics from generated telemetry.

\paragraph*{Workload generation.} Workloads model two labs: LED I-V characterization and RC step response. Each lab is segmented into phases (setup, acquisition, fitting, troubleshooting) with phase-dependent query arrival rates. Arrival rates ($\lambda_{\text{setup}}=0.08$, $\lambda_{\text{acq}}=0.11$, $\lambda_{\text{fit}}=0.14$, $\lambda_{\text{troubleshoot}}=0.09$ requests/minute) were estimated from 45-minute calibration sessions where instructors replayed historical logs and annotated expected hint frequencies. Hint-level transitions within phases follow a Markov model fit to calibration traces (e.g., 0.63 probability of escalating from L1 to L2 after two unanswered attempts). Integrity scenarios (fabricated data flags, range oscillation) are resampled from logs.

\paragraph*{Router execution.} For each synthetic event, the simulator issues \texttt{POST /route/plan} to the production router container, records plan metadata, applies approval and budget logic, and logs telemetry. Freezes persist for 3 turns; boosts apply to the current turn only.

\paragraph*{Conditions.} We swept: embedding providers (off, bge-small-en-v1.5, gte-large-en-v1.5, nomic-embed-text-v1.5), cache TTL (5~s vs.\ 5~min), top-$k$ (1/3/5), and warm vs.\ cold start. The matrix covers 2 labs $\times$ 3 policy levels (P0, P1, P2) $\times$ 2 overlay modes $\times$ 4 seeds, yielding 416 configurations and 208,208 synthetic events. The simulator uses a compact 3-intent canonical bank for ablation convergence. Sensitivity analyses perturbed calibration parameters by $\pm$10--20\%.

\subsection{100-Query Router Replay}
We replayed a 100-query workload through three paths: direct GPT-5 Mini, direct gpt-oss-20b, and EduRouter+Routiium with governed policy.

The workload contains 60 RC-step queries (20 easy, 25 moderate, 15 advanced) and 40 LED I-V queries (15 easy, 15 moderate, 10 advanced). Advanced queries include graduate-level troubleshooting (e.g., ``diagnose ringing in cascaded RC steps with non-ideal op-amps'').

This workload was constructed to align with the 89-entry canonical bank. Consequently, canonical metrics (\chr, \fcr) reflect best-case performance rather than expected behavior on novel queries.

Each replay logged end-to-end latency, routing latency, streaming time-to-first-token, token usage, and canonical match information. Cost calculations used April 2025 list prices: GPT-5 Mini at \$0.25/Mtok input and \$2.00/Mtok output; gpt-oss-20b at \$0.00 (local deployment).

Routing correctness for difficulty bucket $d$ is:
\begin{equation}
c_d = \frac{1}{|Q_d|}\sum_{q \in Q_d} \mathbf{1}[\texttt{tier}(q)=\texttt{preferred\_tier}(d)]
\end{equation}
where $\texttt{preferred\_tier}(d)$ is from canonical metadata.

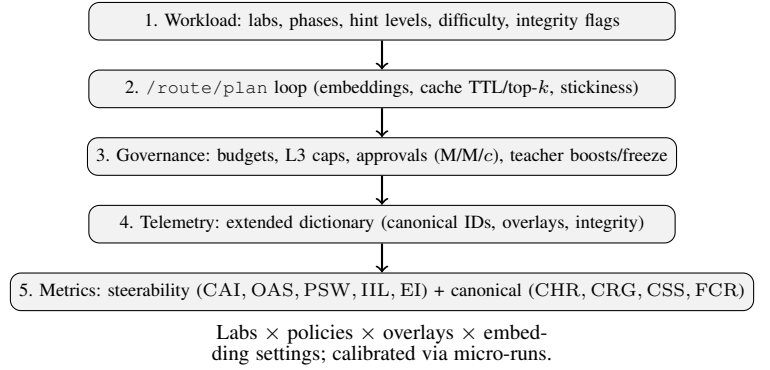
\begin{figure}[t]
\vspace{-0.1in}
\centering
\begin{tikzpicture}[
    font=\scriptsize,
    block/.style={draw, rounded corners, fill=black!5, minimum width=.88\linewidth, minimum height=5mm}
]
\node[block] (workload) at (0,0)      {1. Workload: labs, phases, hint levels, difficulty, integrity flags};
\node[block] (router)   at (0,-0.9)   {2. \texttt{/route/plan} loop (embeddings, cache TTL/top-$k$, stickiness)};
\node[block] (govern)   at (0,-1.8)   {3. Governance: budgets, L3 caps, approvals (M/M/$c$), teacher boosts/freeze};
\node[block] (telemetry) at (0,-2.7)  {4. Telemetry: extended dictionary (canonical IDs, overlays, integrity)};
\node[block] (metrics)  at (0,-3.6)   {5. Metrics: steerability (\(\cai,\oas,\psw,\iil,\ei\)) + canonical (\(\chr,\crg,\css,\fcr\))};
\node[text width=0.9\linewidth, align=center] at (0,-4.3) (notes) {\footnotesize Labs $\times$ policies $\times$ overlays $\times$ embedding settings; calibrated via micro-runs.};
\draw[->, thick] (workload) -- (router);
\draw[->, thick] (router) -- (govern);
\draw[->, thick] (govern) -- (telemetry);
\draw[->, thick] (telemetry) -- (metrics);
\end{tikzpicture}
\caption{Trace-driven simulator pipeline used for synthetic evaluation.}
\label{fig:sim}
\vspace{-0.1in}
\end{figure}

\section{Results}
\label{sec:results}

\subsection{Steerability Metrics (Simulator)}
Table~\ref{tab:governance-metrics} reports steerability metrics across policy conditions.

\begin{table}[t]
    \caption{Steerability metrics by policy (simulator, mean across seeds; \iil\ in turns).}
    \label{tab:governance-metrics}
    \centering
    \begin{tabular}{@{}lccccc@{}}
        \toprule
        Policy & \cai & \oas & \psw & \iil & \ei \\
        \midrule
        P0 (Ungoverned) & 0.90 & 0.69 & 1.45 & 5.20 & 0.59 \\
        P1 (Governed) & 0.98 & 0.84 & 2.75 & 3.20 & 0.56 \\
        P2 (Integrity) & 0.98 & 0.87 & 3.56 & 2.39 & 0.52 \\
        \bottomrule
    \end{tabular}
\end{table}

Under governed policies, \cai\ increased from 0.90 (P0) to 0.98 (P1/P2), indicating closer agreement between observed and target hint distributions. \oas\ increased from 0.69 to 0.87, indicating higher overlay verification rates. \psw\ increased from 1.45 to 3.56 turns, indicating more simulated turns before high-scaffold hints under governed policies.

\ei\ decreased from 0.59 to 0.52, indicating that governed policies concentrated L3 hints among fewer students in the simulation. This is expected given budget limits but represents a tradeoff that instructors would need to consider.

These results characterize simulator behavior under the specified conditions. Whether similar patterns would emerge in classroom deployment depends on factors including accuracy of the student behavior model, which we have not validated.

\subsection{Canonical Metrics (Simulator)}
Table~\ref{tab:canonical-metrics} reports canonical metrics averaged across simulator jobs.

\begin{table}[t]
    \caption{Canonical metrics (simulator, averaged across jobs).}
    \label{tab:canonical-metrics}
    \centering
    \begin{tabular}{@{}lccc@{}}
        \toprule
        Embedding & \chr & \crg & \fcr \\
        \midrule
        gte-large-en-v1.5 & 0.95 & 0.19 & 0.007 \\
        bge-large-en-v1.5 & 0.95 & 0.17 & 0.012 \\
        nomic-embed-text-v1.5 & 0.94 & 0.17 & 0.011 \\
        bge-small-en-v1.5 & 0.93 & 0.15 & 0.015 \\
        FastEmbed (edge) & 0.72 & 0.05 & 0.010 \\
        Embedding off & 0.00 & 0.00 & 0.000 \\
        \bottomrule
    \end{tabular}
\end{table}

Embedding providers achieved \chr\ of 0.93--0.95 on the compact 3-intent bank, with routing gain (\crg) of 0.15--0.19. The FastEmbed edge deployment achieved lower \chr\ (0.72), likely due to encoder mismatch between the quantized ONNX checkpoint and the embeddings used to seed the canonical bank.

Table~\ref{tab:css-metrics} reports canonical stickiness by policy and freeze horizon.

\begin{table}[t]
    \caption{Canonical Stickiness Stability (\css) by policy and freeze horizon.}
    \label{tab:css-metrics}
    \centering
    \begin{tabular}{@{}lcc@{}}
        \toprule
        Policy & Freeze TTL & \css \\
        \midrule
        P0 (Ungoverned) & None & 0.00 \\
        P1 (Governed) & 5~min & 0.49 \\
        P2 (Integrity) & 30~min & 0.60 \\
        \bottomrule
    \end{tabular}
\end{table}

When embedding routing was disabled or freeze keys were flushed, \chr\ and \css\ dropped to zero, confirming that canonical routing requires both library curation and configuration.

\begin{figure}[t]
\vspace{-0.1in}
\centering
\includegraphics[width=.95\linewidth]{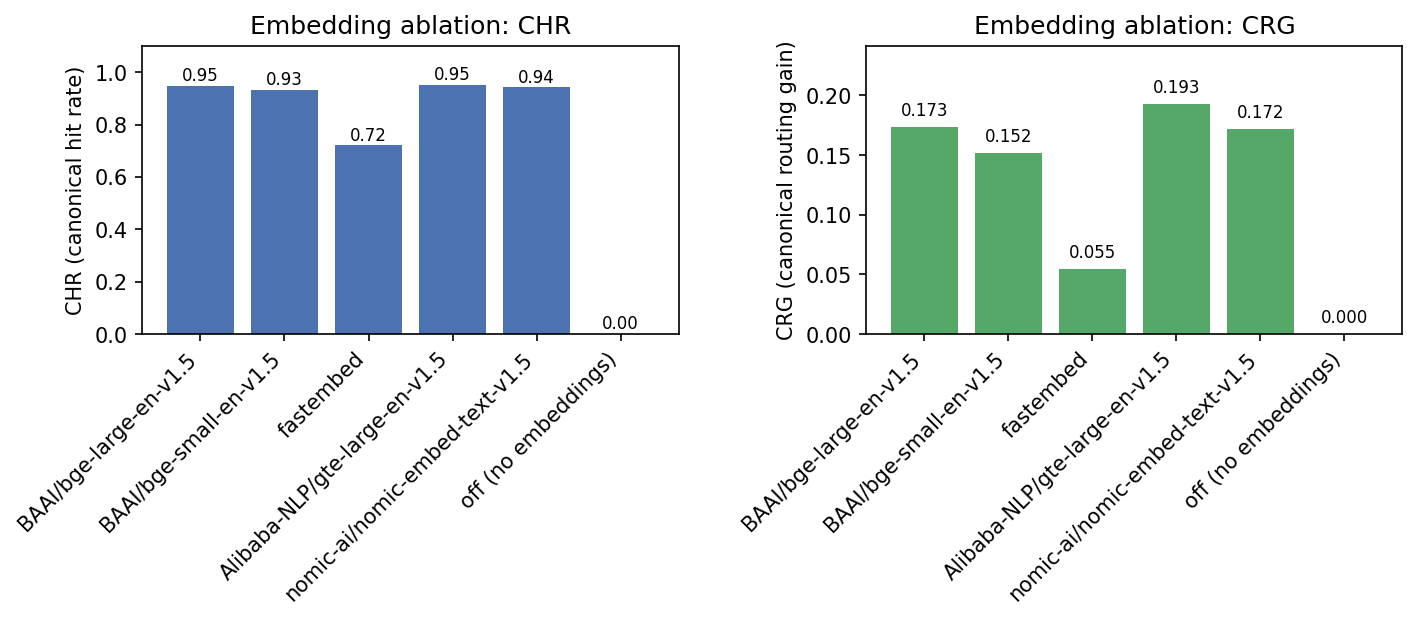}
\caption{Embedding ablation in the simulator. Bars show canonical hit rate (CHR, left) and canonical routing gain (\crg, right) averaged over cache TTLs and warm/cold starts for five embedding configurations---BAAI/bge-small-en-v1.5, BAAI/bge-large-en-v1.5, Alibaba-NLP/gte-large-en-v1.5, nomic-ai/nomic-embed-text-v1.5, and a FastEmbed deployment (ONNX wrapper around BAAI/bge-small-en-v1.5)---plus a no-embeddings baseline. Direct cloud-style integrations of the larger models achieve the strongest CHR and \crg, while the FastEmbed edge deployment trades some canonical fidelity for lower infrastructure cost and dependence on external providers.}
\label{fig:canonical}
\vspace{-0.1in}
\end{figure}

\subsection{Cost and Latency (100-Query Replay)}
\begin{figure*}[t]
    \centering
    \begin{tabular}{cc}
        \includegraphics[width=0.45\textwidth]{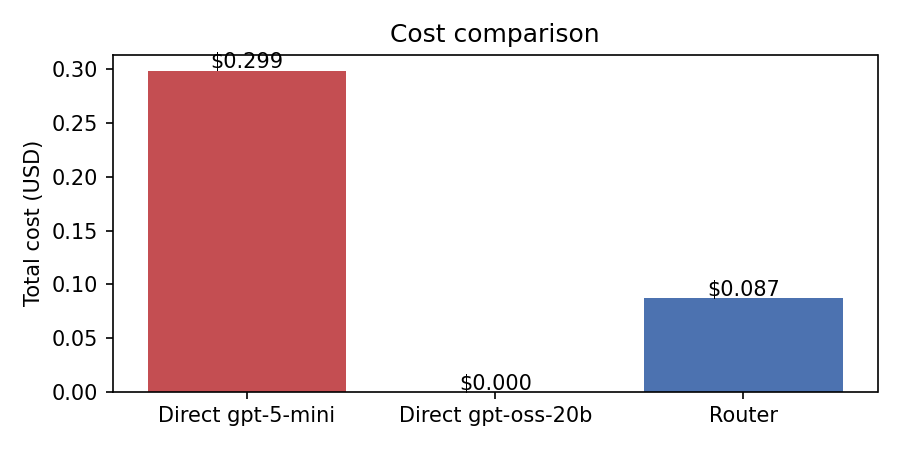} &
        \includegraphics[width=0.45\textwidth]{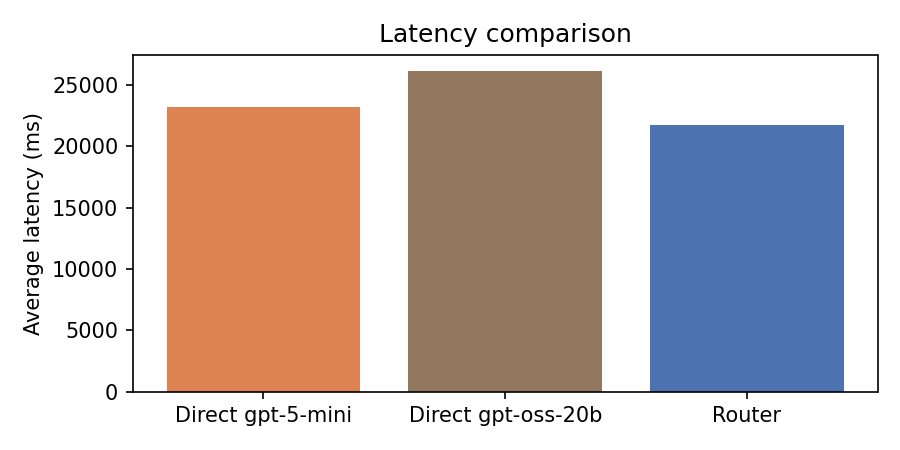} \\
        (a) Cost comparison & (b) Latency comparison \\
        \includegraphics[width=0.45\textwidth]{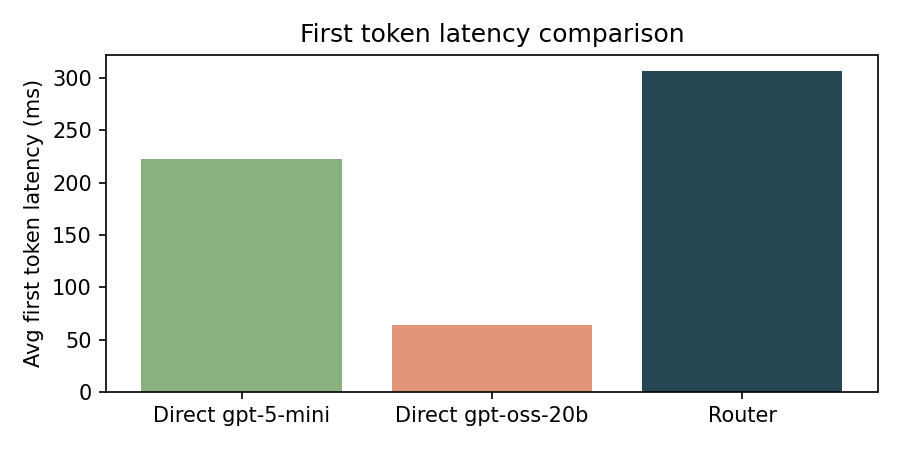} &
        \includegraphics[width=0.45\textwidth]{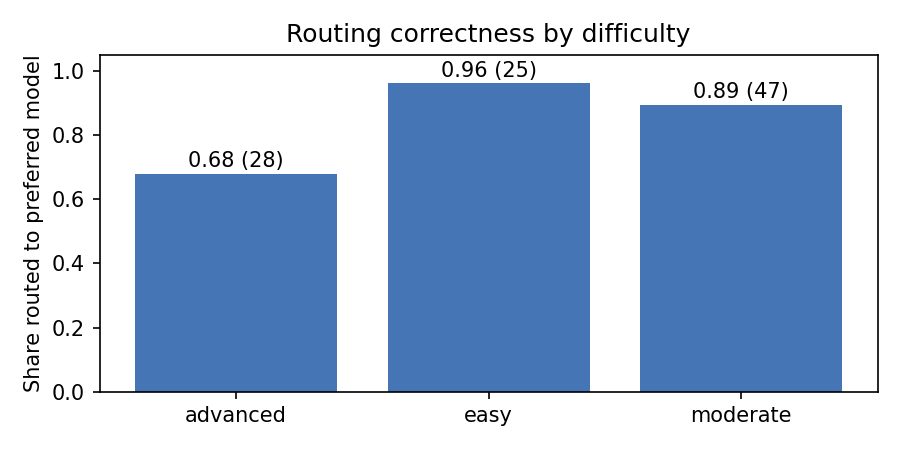} \\
        (c) Time-to-first-token (plan + execution) & (d) Routing correctness by difficulty
    \end{tabular}
    \caption{100-query replay results. (a) Token cost by routing configuration. (b) End-to-end latency. (c) Time-to-first-token decomposed into routing and execution components. (d) Routing correctness by query difficulty.}
    \label{fig:router_eval}
\end{figure*}

Figure~\ref{fig:router_eval} reports results from the 100-query replay.

EduRouter routed 75\% of queries to the local tier, completing the workload for \$0.087 compared to \$0.26 for all-premium routing (66\% reduction). This cost difference reflects the specific workload and pricing; generalization to other workloads or pricing structures would require additional evaluation.

Direct GPT-5 Mini latency averaged 23.2~s with 222~ms time-to-first-token (TTFT). The local tier averaged 26.1~s latency with 64~ms TTFT. Routed execution averaged 21.7~s latency with 305~ms aggregate TTFT (including 85--90~ms routing overhead).

\subsection{Routing Correctness (100-Query Replay)}
Routing correctness by difficulty bucket was: easy 0.96, moderate 0.89, advanced 0.68. The lower correctness for advanced queries reflects incomplete canonical coverage for graduate-level troubleshooting scenarios; the current bank preferentially routes these to the local tier when GPT-5 Mini might be more appropriate.

Canonical metrics for this workload were \chr\ = 1.0 and \fcr\ = 0.0. These values reflect the workload construction (aligned to the canonical bank) rather than expected performance on novel queries.

\begin{table}[t]
\caption{Telemetry feature dictionary (excerpt).}
\label{tab:telemetry}
\centering
\begin{tabular}{ll}
\toprule
\textbf{Field} & \textbf{Meaning} \\
\midrule
session\_id & Anonymous session identifier \\
lab\_id & Micro-lab (\texttt{led\_iv}, \texttt{rc\_step}) \\
policy & \texttt{P0}, \texttt{P1}, or \texttt{P2} \\
step\_id & Current lab step key \\
hint\_req / hint\_granted & Request/grant events + level (L0--L3) \\
justification\_len & Characters in student justification \\
model & Routed model identifier \\
overlay\_fingerprint & Active overlay stack hash \\
canonical\_ids / scores & Canonical matches + similarity score \\
canonical\_reason & Text explaining canonical bias \\
tokens & Prompt+completion tokens (per exchange) \\
latency\_ms & Round-trip latency \\
est\_cost\_micro & Plan cost estimate in micro-dollars \\
teacher\_boost & Boolean; teacher forced stronger model \\
approval\_id / wait\_ms & Approval decision + queue latency \\
privacy\_mode & \texttt{full}, \texttt{hashed}, or \texttt{off} embeddings \\
scpi\_retries & Instrument I/O retry count \\
range\_changes & Number of range/scale flips \\
integrity\_flag & Fabricated/at-risk telemetry indicator \\
step\_pass & Auto acceptance test pass (Y/N) \\
rubric\_score\_blind & Synthetic rubric-like score (0--100) \\
\bottomrule
\end{tabular}
\end{table}

\begin{table}[t]
\caption{Policy targets used in simulation scenarios.}
\label{tab:policy}
\centering
\begin{tabular}{@{}p{0.28\columnwidth}p{0.62\columnwidth}@{}}
\toprule
\textbf{Knob} & \textbf{Target / Rationale} \\
\midrule
Hint budgets & \$5 total / lab; \texttt{L3\_max}=2 to enforce \psw goals \\
Approvals & Two TAs (M/M/2) with p95 $<$ 60~s decision latency \\
Overlay personas & Default ``Socratic troubleshooting''; teacher can swap to ``diagnostic'' overlay in P2 \\
Boost/freeze & One-click teacher boost to premium tier; freeze TTL default 5~min (extendable to 30~min) \\
Integrity throttle & Block assistance when integrity\_flag triggers for 3 consecutive turns \\
Canonical embedding & Enabled in P1/P2 via \texttt{fastembed}; hashed fallback when privacy\_mode=\texttt{hashed} \\
Queue discipline & FIFO approvals with justification gating for L3 requests \\
Telemetry retention & 30 days for analytics; canonical IDs stored with HMAC for audits \\
\bottomrule
\end{tabular}
\end{table}

\begin{table}[t]
\caption{Governance and canonical metric glossary.}
\label{tab:metrics}
\centering
\begin{tabular}{ll}
\toprule
\textbf{Metric} & \textbf{Interpretation} \\
\midrule
\cai & Challenge alignment (hint mix vs.\ targets) \\
\oas & Overlay adherence (no-leakage rate) \\
\psw & Productive-struggle window (steps before L2/L3) \\
\iil & Instructor-influence latency (turns to take effect) \\
\ei & Equity index ($1-\text{Gini}$ on L3 allocation) \\
\chr & Canonical hit rate (fraction with canonical match) \\
\crg & Canonical routing gain (simulated cost/latency lift) \\
\css & Canonical stickiness stability (tier retention rate) \\
\fcr & False-canonical rate (fallbacks after canonical pick) \\
Autonomy & Autonomy gain (trend in L3 share over time) \\
TTFT & Time-to-first-token (plan + exec first chunk) \\
RC$_\text{diff}$ & Routing correctness by difficulty bucket \\
\bottomrule
\end{tabular}
\end{table}

\section{Limitations and Threats to Validity}
\label{sec:threats}

Several limitations constrain interpretation of these results.

Metrics computed from simulation depend on the fidelity of the student behavior model. Arrival rates and hint-escalation probabilities were estimated from limited calibration sessions (45 minutes per lab) and may not represent actual student behavior. The Markov model for hint escalation assumes state transitions depend only on current state, which may oversimplify actual student decision-making.

The 100-query workload was constructed to align with the canonical bank, so \chr\ = 1.0 represents a ceiling rather than expected performance. Performance on novel queries not covered by the bank would be lower and would depend on fallback routing quality. The lower routing correctness for advanced queries (0.68) indicates current canonical coverage is insufficient for graduate-level content.

Results are specific to two electronics labs. Whether similar patterns would emerge for other lab types (software, mechanical, conceptual coursework) is unknown. Cost savings depend on the specific pricing structure and workload characteristics; different model pricing or query distributions would yield different results.

We have not evaluated whether instructors find the governance mechanisms useful or whether the curation effort (estimated 2--3 hours per term) is acceptable. The console interface was designed based on anticipated needs rather than user studies. We do not claim effects on student learning; the metrics characterize router behavior, not pedagogical effectiveness. Whether higher \psw\ corresponds to better learning outcomes is an empirical question requiring classroom studies with appropriate controls.

The simulation showed \ei\ decreasing under governed policies, indicating concentrated L3 allocation. Whether this concentration is educationally harmful or appropriate (e.g., students with greater need receiving more assistance) depends on context we have not examined. Additionally, embedding-based routing requires computing query representations; when these computations occur off-campus, privacy implications arise. Biases in canonical entries or overlay prompts could systematically affect assistance quality for some students. We include privacy modes and hashed fallbacks but have not evaluated their effectiveness.

\section{Conclusion}
We described a system for routing LLM-based lab assistance that implements instructor-configurable policies on hint levels, approvals, and budgets, with embedding-based question matching and detailed audit logging. Simulation-based evaluation indicates the system can enforce policies under synthetic workloads, and the 100-query replay demonstrates cost reduction through tiered model routing.

These results characterize system behavior, not educational effectiveness. Next steps include classroom deployment to study whether the governance mechanisms reduce instructor workload and whether policy compliance correlates with learning outcomes. Additional work is needed to validate the student behavior model against real classroom data, expand canonical coverage for advanced queries, and study instructor adoption of the curation and monitoring workflows.

All source code and artifacts are available at: \texttt{https://github.com/labiium/edurouter} (EduRouter) and \\ \texttt{https://github.com/labiium/routiium} (Routiium Gateway). We release Routiium source code, EduRouter with schema version 1.1, canonical bank tooling, simulator configurations, and analysis notebooks. The simulator bundle includes aggregated CSV summaries and an execution manifest with policy hashes and seeds. The router replay bundle provides JSONL traces and derived summaries.

\bibliographystyle{IEEEtran}
\bibliography{paper}
\end{document}